# Realistic Haptic Rendering of Interacting Deformable Objects in Virtual Environments

Christian Duriez, *Student Member*, *IEEE*, Frédéric Dubois,
Abderrahmane Kheddar, *Member*, *IEEE*, and Claude Andriot

**Abstract**—A new computer haptics algorithm to be used in general interactive manipulations of deformable virtual objects is presented. In multimodal interactive simulations, haptic feedback computation often comes from contact forces. Subsequently, the fidelity of haptic rendering depends significantly on contact space modeling. Contact and friction laws between deformable models are often simplified in up to date methods. They do not allow a "realistic" rendering of the subtleties of contact space physical phenomena (such as slip and stick effects due to friction or mechanical coupling between contacts). In this paper, we use Signorini's contact law and Coulomb's friction law as a computer haptics basis. Real-time performance is made possible thanks to a linearization of the behavior in the contact space, formulated as the so-called Delassus operator, and iteratively solved by a Gauss-Seidel type algorithm. Dynamic deformation uses corotational global formulation to obtain the Delassus operator in which the mass and stiffness ratio are dissociated from the simulation time step. This last point is crucial to keep stable haptic feedback. This global approach has been packaged, implemented, and tested. Stable and realistic 6D haptic feedback is demonstrated through a clipping task experiment.

---

♦

---

## 1 INTRODUCTION

RECENT advances in virtual reality enable haptic exploration of virtual objects through various haptic interface technologies [1]. Although some deformation algorithms of the virtual objects are modeled and solved in real-time using acceleration methods (Section 1), the majority of these methods are based on the hypothesis of single point interaction [2], [3]. With this assumption, a precomputation stage is required to enable real-time interactivity. Unfortunately, extending these methods to complex multiobject interactions cannot be easily achieved. In addition, other common contact methods such as the penalty or imposed motion (see Section 2) do not adequately solve multiple deformable object scenes. In this work, we focus on the computation of the contact forces (Section 3) between deformable objects including friction (Section 4), resulting in realistic haptic feedback.

The majority of contact methods involving friction between virtual objects in haptics are based on Coulomb's law. This nonlinear law describes two states on the tangential contact space: the stick and the slip. This law is difficult to solve correctly in a multicontact context. Between deformable objects, previous methods often compute Coulomb's law explicitly from the data of the previous time step. This leads to a nice visual behavior, but the haptic feedback incurs drifts mainly during the stick phase.

In the rigid body context, approximation strategies can be used to allow implicit computation, such as $k$-sided pyramids. We prove in this paper that it is not the fastest strategy. A central contribution of this paper is the Gauss-Seidel iterative algorithm adapted to multicontact and the friction case does not necessitate any approximation on Coulomb's law.

To add these new components, the kinesthetic part is incorporated within the physically-based engine which computes interaction forces and behavior of several virtual objects based on physical contact models. A manipulated object, attached in some manner (see Section 7) to the haptic display, is just part of the surrounding environment. Since forces are already computed within the simulation engine, obtaining haptic feedback forces can be easily extracted.

The realism of the contact models is a crucial issue in haptics. If the contact space model is not correct, the transparency, i.e., the rendering fidelity, will be affected whatever the sophistication of the haptic display technology. Indeed, in most applications [4], [5], haptic rendering reflects the interactions due to contacts. They may be numerous, and the topology of the contact space complex. The next section introduces the difficult issue of calculating contact and friction between deformable bodies.

## 2 CONTEXT

### 2.1 Deformation Models Used in Real-Time Simulations

It is well-known that elastic objects have infinite degrees of freedom and the deformation behavior equations are impossible to solve analytically. Many authors have proposed different techniques to speed up deformation computations based on the finite element methods (FEM)

- *C. Duriez is with CIMIT Simulation Group, 65 Lansdowne Street, Suite 142, Cambridge, MA 02139. E-mail: christian.duriez@gmail.com.*
- *F. Dubois is with LMGC-UMR 5508, Université MONTPELLIER II, CC 048 Place Eugène Bataillon, 34095 Montpellier cedex 5, France. E-mail: dubois@lmgc.univ-montp2.fr.*
- *A. Kheddar is with the AIST/IS-CNRS/STIC Joint Japanese-French Robotics Laboratory JRL, AIST Tsukuba Central 2, Umezono 1-1-1, Tsukuba 305-8568, Japan. E-mail: abderrahmane.kheddar@aist.go.jp.*
- *C. Andriot is with CEA/LIST-SCRI, Route du Panorama, 92165 Fontenay-aux-Roses, France. E-mail: claude.andriot@cea.fr.*

[6], [7], [8]. Virtual objects are considered to be already meshed and initialized with all related physical parameters such as intrinsic mass and internal forces known in analytic or numerical form. FEM methods are preferred to classical discrete mechanical elements (DME) (gathering mass-spring and particle based methods), but both classes of methods lead eventually to a similar mathematical matrix formulation:

$$M\ddot{\mathbf{u}}_t + D\dot{\mathbf{u}}_t + K(\mathbf{u}_t) = \mathbf{f}_t, \qquad (1)$$

where $M$ and $D$ are, respectively, FEM or DME mass and damping matrices, $K(\mathbf{u}_t)$ represents internal forces, $\mathbf{f}_t$ is the current applied force field equivalently distributed to each node of the object's mesh, and $\mathbf{u}_t$ are the current nodes' displacements. Different methods are then used to linearize the problem in order to achieve real-time computation:

- For small displacements and linear elasticity: $K(\mathbf{u}_t) \simeq K\mathbf{u}_t$.
- The internal forces may be explicitly formulated: $K(\mathbf{u}_t) \simeq K(\mathbf{u}_{t-1}, \dot{\mathbf{u}}_{t-1}, \ddot{\mathbf{u}}_{t-1})$. In this case, internal forces are shifted to the right-hand side of the equation, having $K(\mathbf{u}_t) \simeq \mathbf{f}^K_{t-1}$, and subtracted from applied external forces $\mathbf{f}_t$.
- Internal forces may be linearized: $K(\mathbf{u}_t) \simeq K'(\mathbf{u}_{t-1}, \dot{\mathbf{u}}_{t-1}, \ddot{\mathbf{u}}_{t-1})\mathbf{u}_t$.

Thus, applying order 1 or 2 numerical integration to (1) leads to:

$$\widetilde{K}\mathbf{u}_t = \widetilde{\mathbf{f}}_{(t,t-1,t-2)}. \qquad (2)$$

$\widetilde{K}$ is analogous to a stiffness that depends on: 1) the intrinsic properties of the object, 2) the numerical integration method, and 3) the way $K(\mathbf{u})$ is linearized (see [9] for details).

## 2.2 Contact Solving between Deformable Bodies

Consider several objects[1] moving randomly and coming into contact with each other. Rendering must guarantee realistic deformation while preventing object interpenetration. We will develop a theory for the case of a pair of objects in contact.[2]

Collision (or proximity) detection (CD) allows characterizing the surface area where object could be "potentially" in contact. This paper does not make new contributions in the area of CD. Interested readers may refer to [10], [11] for a review. After CD, we can build the contact space (i.e., colliding (or proximity) spots to each of which a normal is associated, see [12], for instance).

When the contact space is defined, collisions need to be resolved. The response is an update of nodes' positions inducing visual deformation. As stated before, the deformation behavior can be linearized to:

$$\widetilde{K}_i \mathbf{u}_i = \mathbf{f}_i, \qquad (3)$$

where the index $i$ is the numerical label of an object. At the start of the contact process, $\mathbf{u}_i$ and $\mathbf{f}_i$ are unknown.

When contacts have coupling and friction, one cannot presuppose the value of the forces or the behavior of the node displacements. Therefore, there is one equation of motion for two vector unknowns and a contact model which is not smooth. To avoid dealing with the contact model as in most previous approaches, a hypothesis is made about one of these two unknowns. The first approach is that the force value is directly related to an interpenetration measure: This is the penalty method. The second approach, called the imposed motion method, freezes motion along the normal to all contacts.

## 2.3 Penalty Methods

Largely used for their simplicity, they require a penetration measurement (e.g., depth, volume, etc.) which is given by CD. If $\delta$ is the scalar norm of the penetration,[3] then $\mathbf{f}_i = -k_i\delta - b_i\dot{\delta}$. Here, $k_i$ and $b_i$ are fixed arbitrary values (sometimes, $b_i = 0$) that may be adapted during the computation process [14], [13], [15].

Stability problems could arise when the integration method is explicit. The penetration measurements during CD give the values of the contact forces based on the penalty law. Deformations then use these forces. In this case, the contact solving process is straightforward, but contact forces will only depend on the contacts' geometry and on the arbitrary choice of the penalty factor. When the method is implicit, the penalty laws are added to (3). The process leads to a comparatively better solution if (and only if) the penalty factor has a large value (a much higher value than the elasticity modulus of the objects). However, this leads to a nonlinear[4] stiff problem.

## 2.4 Imposed Motion

In imposed motion methods, contacts are considered as "bilateral" constraints. The imposed motion of the contact nodes are calculated using lagrangian multipliers [16]. This fixes $\mathbf{u}_i$, in the contact space, according to a master-slave scheme (the stiffer object is the master). The method may result "sticky" effects during relative objects motion when a contact force is badly oriented from the removal of an associated contact from the list in order to keep unilateral constraints [17]. This works well when the contact topology changes are smooth, such as when the bodies are very soft and collision velocities are low.

These methods can be very fast and relatively accurate in some cases, but they cannot be easily generalized to contact between all type of deformable objects. To the best of our knowledge, they are always applied in frictionless contexts. Moreover, imposing motion in the tangential motion space is very problematic, particularly when the adherence is supposed to be weak for nonrough surfaces, which is likely to be the case in many applications.

## 2.5 Discussion

In the context of haptics, it appears that the interaction of multiple deformable objects has not been adequately solved. Based on the literature in this area and our experience in haptics, we make the following observations:

---

1. From now on, object(s) are considered to be virtual and deformable.
2. This is the most common situation in haptics, but we clearly demonstrate an extension to $n$-bodies.

3. In the multicontact case, values given for each contact are composed into a vector noted $\delta$.
4. The simplest penalty law is:
$$f = -k\delta \quad \text{if} \quad \delta < 0$$
$$f = 0 \quad \text{if} \quad \delta \geq 0.$$
In this case, the function $f$ with respect to $\delta$ is piecewise-linear. When only one contact point is considered, it is easy to consider separately the two linear problems (if $\delta < 0$ or if $\delta \geq 0$) But, for two contacts, one obtains four different linear cases, with three contacts eight different linear cases, for 20 contacts, $2^{20}$ linear cases. So, generally, in multicontact case, the problem is considered and treated as a nonlinear problem.

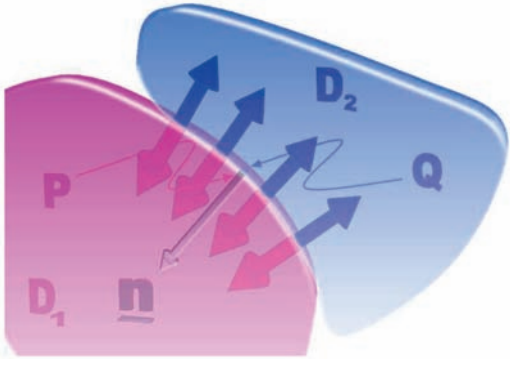

Fig. 1. Contact between two deformable bodies.

- Contact models greatly influence the quality of haptic feedback.
- Haptics output from real-time deformation algorithms is based on simplified contact laws.
- 3D Coulomb's friction model (or resolution in case of multicontact) is simplified in current haptic simulations.

Our work is the first to consider the problem of haptic manipulation of several deformable objects with no-penetration constraints and friction as described below.

## 3 MODELING CONTACT FROM THE SIGNORINI'S LAW

Our first contribution is in using Signorini's law to resolve the contacts between two bodies in a real-time multimodal interactive simulation context. Signorini's law is known in continuous media mechanics and is used in offline simulation of rigid or deformable objects [18], [19]. However, to the best of our knowledge, its real-time resolution for deformable objects and computer haptics has never been investigated. The next section briefly describes its formulation for the frictionless case (a detailed discussion can be found in [20]).

### 3.1 Formulation in Continuous Media Mechanics

We are using Signorini's law to resolve the contact between two bodies labeled $D_1$ and $D_2$. To each particle $P$ of $D_1$ *potentially* in contact with $D_2$, we associate a neighboring particle $Q$ of $D_2$ to test the contact between $D_1$ and $D_2$ (see Fig. 1). The direction of $\overrightarrow{QP}$ is given by $\underline{n}$.

Consider the unknown $\sigma_n^{(1)}(P)$ (the stress exerted on $D_1$ in $P$). We have:

$$\sigma_n^{(1)}(P) + \sigma_n^{(2)}(Q) = 0. \quad (4)$$

The normal $\underline{n}$, chosen arbitrarily,[5] is directed toward the inside of $D_1$. The gap between the two objects at $P$ is:

$$\delta_n(P) = \overline{QP} \,\overline{\otimes}\, \underline{n}. \quad (5)$$

(The contracting product $\overline{\otimes}$ between two vectors is a dot product).

---

[5]. We could have used the direction $-\underline{n}$ of $\overrightarrow{PQ}$. By choosing the direction $\underline{n}$, the problem is solved by considering the unknown forces applied at points $P$ on $D_1$. We could get the same solution by using the opposite direction and taking the unknown forces applied at point $Q$ on $D_2$.

The Signorini contact model indicates that there is a complementarity relation[6] between this gap $\delta_n(P)$ and the Cauchy stress $\sigma_n^{(1)}(P)$, that is:

$$0 \leq \delta_n(P) \perp \sigma_n^{(1)}(P) \geq 0. \quad (6)$$

This model has several physical justifications:

- $\delta_n(P) \geq 0$ guarantees noninterference.
- The pressure exerted by $D_2$ on $D_1$ is inevitably directed toward object $D_1$, i.e., $\sigma_n^{(1)}(P) \geq 0$.
- If the contact between objects at $P$ is active, $\delta_n(P) = 0$ and $D_2$ exerts a pressure on $D_1$ at point $P$. Otherwise, $\delta_n(P) > 0$ and the stress exerted by $D_2$ at $P$ is null.

Signorini's law does not give any indication of the tangential constraints in the contact space. Coulomb's law will be added in Section 4.

### 3.2 Collision/Proximity Detection Issues

We assume that a collision/proximity detection algorithm identifies $m$ potential contacts between a pair of bodies $D_1$ and $D_2$. For every contact, one normal vector is provided.

As the surface of bodies is meshed with triangles, most of the contacts will appear as the canonical collision of two triangles (i.e., a collision of a vertex of one triangle with the face of the other, or a collision between two edges. See the two cases at the top of Fig. 3.).

However, other cases are possible. As Fig. 3 shows, in all cases, several contact points may describe the contact. For each contact point, we need collision (or proximity) detection to provide:

- two contact points $P$ and $Q$,
- their barycentric position within their triangle, and
- (eventually) the contact normal $\vec{n}$. If not provided, $\vec{n}$ is set to the initial direction of $\overrightarrow{QP}$.

This data can be provided by a number of current CD algorithms adapted to deformable bodies. However, the nature of the shapes (nonconvexity, nonsmoothness, fast variation of surfaces) may influence their performance. Since this is the input to our algorithm, the data accuracy from CD will influence our results. However, we make no specific assumption about CD. As long as a collision/proximity detection can provide the information needed, our algorithm works robustly.

If CD misses some intersections, interpenetration between deformable bodies could be observed. However, if the missed intersections are reported in the subsequent time steps, a correct configuration can be restored. If the correct contact configuration is restored quickly and motions are not too fast, such events will not induce instabilities.

The goal of this paper is to clearly provide a solution to multifrictional-contact solving between deformable models, assuming efficient CD. The following section describes how information provided by CD is used in an FEM formulation.

### 3.3 Discrete Formulation Using Linear Finite Elements

Let $\Psi$ be the interpolation functions used in FEM. Indeed, the position of a point inside an element depends on the position of the nodes and on the interpolation functions.

---

[6]. Complementarity is noted $\perp$; this relation states that one of the two values $\delta_n(P)$ or $\sigma_n^{(1)}(P)$ must be null.

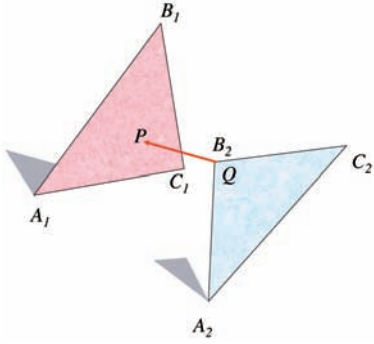

Fig. 2. Each contact connects two points $P$ and $Q$ that are interpolated to the nodes of the mesh, respectively, $A_1$, $B_1$, $C_1$ and $A_2$, $B_2$, $C_2$.

This work is based on linear interpolation functions using tetrahedrons with four nodes. Although not very precise, when combined with the contact model, they allow the use of an equivalent formulation of Signorini's problem in stress or in force (see [20] for details).

Thus, using $f_n$, the contact force on each potential collision point output from CD, we can write:

$$0 \leq \delta_n(P) \perp f_n(P) \geq 0 \Leftrightarrow 0 \leq \delta_n(P) \perp \sigma_n^{(1)}(P) \geq 0. \quad (7)$$

In FEM construction, each contact point is interpolated from the nodes of the element to which it belongs. The points are necessarily on the surface of the objects on triangles. Let $(A_1 B_1 C_1)$ (Fig. 2) be the support triangle for $P$:

$$U_P = \sum_{\alpha = \{A_1, B_1, C_1\}} \Psi_\alpha(P) U_1(\alpha). \quad (8)$$

In the same way, for $(A_2 B_2 C_2)$ and $Q$:

$$U_Q = \sum_{\alpha = \{A_2, B_2, C_2\}} \Psi_\alpha(Q) U_2(\alpha). \quad (9)$$

$U_1$ and $U_2$ are the node displacements of $D_1$ and $D_2$, respectively. $U_1(\alpha)$ is the 3D displacement of $D_1$'s node $\alpha$.

As the interpolation functions are linear inside the elements, one can equivalently interpolate the unknown contact force at $P$, named $f_P$.

$$\vec{F}_1(\alpha) = \Psi_\alpha(P) \vec{f}_P \text{ with } \alpha = \{A_1, B_1, C_1\}. \quad (10)$$

And, similarly:

$$\vec{F}_2(\alpha) = \Psi_\alpha(Q) \vec{f}_Q \text{ with } \alpha = \{A_2, B_2, C_2\}. \quad (11)$$

$F_1$ and $F_2$ are the force vectors in the space of the mesh nodes of $D_1$ and $D_2$, respectively. And, $\vec{F}_1(\alpha)$ is the force applied to $D_1$'s node $\alpha$.

With two objects, there is necessarily an arbitrary choice for the direction of the unknown force. In the direction $\underline{n}$, $f_P$ is positive. This corresponds to the force that the object $D_2$ exerts on $D_1$:

$$\vec{f}_P = -\vec{f}_Q = \vec{n} f_n. \quad (12)$$

In the contact space, the force $f_n$ is scalar. It will be projected onto the contact normal.

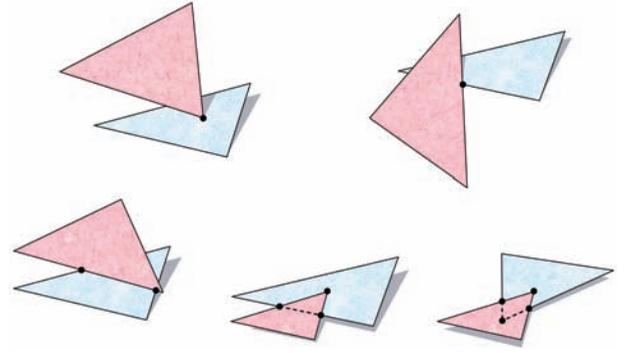

Fig. 3. A collision between two triangles leads to several cases in terms of contact point number. However, in all of these cases, it is possible to have sufficient contact points to describe Signorini's law correctly in terms of the contact force exerted on these points.

### 3.4 Linear Complementarity Problem (LCP) Formulation

To linearize Signorini's problem, the contact space is frozen during the current time step such that both the gap $\delta_n(P)$ and force $f_n$ on each contact can be considered as strictly positive scalars.

With CD, one may compute the value of each gap for the current time step if no contact forces are applied to the object. This is called the free motion, and the corresponding gap, projected along the contact's normal, is denoted $\delta_n^{\text{free}}(P)$.

Deformation displacements $U_P$ and $U_Q$ are considered between the free motion and the constrained motion (after resolution of Signorini's problem and the integration of the contact forces):

$$\delta_n(P) = n^T (U_P - U_Q) + \delta_n^{\text{free}}(P). \quad (13)$$

By stacking relations (8), (9), and (13) for each contact, one can write matrices $H_1$ and $H_2$ such that:

$$\delta_n = [H_1][U_1] - [H_2][U_2] + \delta_n^{\text{free}}. \quad (14)$$

$U_1$ and $U_2$ are the displacements of the nodes involved in the contact. In the same way, by stacking all the forces relations ((10) and (11)), we have:

$$\begin{aligned} [F_1] &= [H_1]^T f_n \\ [F_2] &= -[H_2]^T f_n. \end{aligned} \quad (15)$$

We now need to formulate a linear relation linking the contact forces to the relative positions in contact space. As stated in (3), it is possible to find a matrix $C_i = (\widetilde{K}_i)^{-1}$, analogous to a compliance, also named capacitance in [17]. In linear deformation cases, or when the stiffness may be explicitly considered, the compliance may be preprocessed. If the stiffness is linearized at every time step, the compliance matrix can be condensed to only only nodes involved in the contact.[7] In this case, the compliance is not computed fast.

Having a mechanical compliance, a linear relation links the constrained node positions $P$ and external contact forces

---

7. If one node $\alpha$ is not involved in a contact, the node's force $F_\alpha$ is null in (16). In order to build the LCP, the motion of $\alpha$ is not necessary. Then, a condensate formulation of (16), excluding the node $\alpha$, can be used.

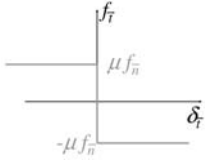

Fig. 4. Coulomb's friction law.

$F$ at $P$. Additionally, positions obtained from free motion $P^{\text{free}}$ appear in the same expression, that is:

$$P = C_i F + P^{\text{free}}. \tag{16}$$

For a pair of objects, the expected linear relation is:

$$\delta_n = ([H_1]C_1[H_1]^T + [H_2]C_2[H_2]^T)f_n + \delta_n^{\text{free}}. \tag{17}$$

If multiple objects are in contact in the same time step, we can use the same formulation. The whole of the objects linked by one or more contact(s) form a contact group. Then, for every contact group, this formulation leads to the following LCP:

$$\begin{cases} f_n \geq 0 \\ \delta_n = [\Sigma_i H_i C_i H_i^T]f_n + \delta_n^{\text{free}} \geq 0 \\ f_n \perp \delta_n. \end{cases} \tag{18}$$

The matrix $[\Sigma_i H_i C_i H_i^T]$ is also known as the Delassus operator [21] and, from now on, it is labeled $[W]$. The obtained LCP in (18) may be solved using several methods, see [22]. In [20], simulation results in the frictionless case are presented. A more truthful simulation needs to take into account another phenomenon: static and dynamic friction.

## 4 COULOMB'S FRICTION LAW

Coulomb's friction law (see Fig. 4) describes the macroscopic behavior in tangent contact space. To keep homogeneity with Signorini's law, an integrated description of Coulomb's law is used. In this law, the tangential gap measured in the contact space is the integrated value of the tangential velocity:

$$\begin{aligned} \delta_{\vec{t}} = \vec{0} &\Rightarrow \|f_{\vec{t}}\| < \mu \|f_{\vec{n}}\| \text{ (stick)} \\ \delta_{\vec{t}} \neq \vec{0} &\Rightarrow f_{\vec{t}} = -\mu \|f_{\vec{n}}\| \frac{\delta_{\vec{t}}}{\|\delta_{\vec{t}}\|} \text{ (slip)}. \end{aligned} \tag{19}$$

The integration of this model with interactive haptic simulations is a very challenging issue. Mathematical difficulties are described first, followed by a discussion of the previously proposed $k$-sided pyramids method that uses approximation of friction cones. We will not use this method because of its asymptotic complexity.

### 4.1 Tangential Nonlinearity

In 3D motion, nonlinearity comes from the tangential direction of the friction force. Consider the slipping motion (dynamic friction) of a single contact point. In Coulomb's law, the direction of the friction force must be in the direction of the tangential motion. But, this motion is unknown and the friction force is too. The only available equations are:

- the linearized system mechanical behavior, along two tangential directions, $\vec{t}_1$ and $\vec{t}_2$:

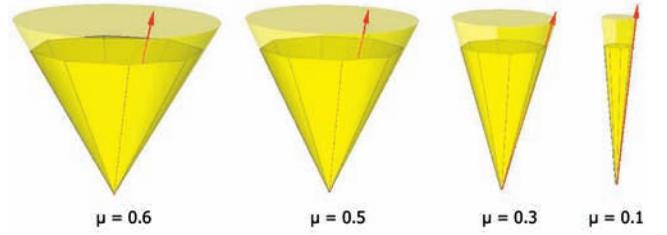

Fig. 5. Approximation of the friction cone by 8-sided pyramids. The hollow cone shows the actual friction model, and the red arrow shows contact forces (static in the first two cases and dynamic in the last ones).

$$\begin{bmatrix} \delta_{\vec{n}} \\ (\delta_{\vec{t}_{1,2}}) \end{bmatrix} = \begin{bmatrix} W_{nn} & (W_{nt})_{(1\times 2)} \\ (W_{tn})_{(2\times 1)} & [W_{tt}]_{(2\times 2)} \end{bmatrix} \begin{bmatrix} f_{\vec{n}} \\ (f_{\vec{t}_{1,2}}) \end{bmatrix} + \begin{bmatrix} \delta_{\vec{n}}^{\text{free}} \\ (\delta_{\vec{t}_{1,2}}^{\text{free}}) \end{bmatrix}. \tag{20}$$

- Coulomb's friction law, in case of dynamic friction, which is a nonlinear relation along the unknown tangential direction of the motion:

$$f_{\vec{t}} = -\mu \|f_{\vec{n}}\| \frac{\delta_{\vec{t}}}{\|\delta_{\vec{t}}\|}. \tag{21}$$

It is not possible to separate the Coulomb's law computations from contact calculations. Each contact's force can modify the state of the other contact spots through the tangent space which is also coupled, locally, to the normal space. In case of a single contact, fast and very precise solutions have been proposed, see [3], for instance. But, in the general multicontact case, since the dynamic friction is nonlinear, it cannot be directly expressed in an LCP form. Thus, approximation of Coulomb's law has been proposed.

### 4.2 $k$-Sided Pyramids Approximation

In rigid body dynamics, well-known approaches formulated contact forces resolution as an LCP [23]. This is made to solve the forces in implicit or semi-implicit schemes. For multicontact interactive simulations with dry friction, a solution often proposed (see [24] or [25]) is to approximate Coulomb's friction cone by a polyhedral one, Fig. 5. This allows the keeping of an LCP formulation.

As the LCP matrix is copositive, Lemke's algorithm allows a solution to be found. The polygonal approximation to the friction cone is written:

$$\widehat{FC}(P) = \text{cone}\{\vec{n} + \vec{t}_r \mid r = 1\ldots k\}, \tag{22}$$

where $k$ is the number of cone faces (triangles). Adapting this formulation to our notation leads to the following formulation:

$$\begin{cases} 0 \leq \delta_{\vec{n}} \perp f_{\vec{n}} \geq 0 \\ 0 \leq \lambda + \delta_{\vec{t}_1} \perp f_{\vec{t}_1} \geq 0 \\ \quad\vdots \\ 0 \leq \lambda + \delta_{\vec{t}_k} \perp f_{\vec{t}_r} \geq 0 \\ 0 \leq \mu f_{\vec{n}} - \sum_{r=1}^{k} f_{\vec{t}_r} \perp \lambda \geq 0. \end{cases} \tag{23}$$

Note that $\lambda$ is not a physical quantity, although it usually represents the sliding displacement at the contact, see [25].

The methodology used to set the Delassus operator $[W]$ along a contact normal can be applied to write this operator along normal and tangential components. That is, for one contact:

$$0 \leq \begin{bmatrix} f_n \\ f_{t_1} \\ \vdots \\ f_{t_k} \\ \lambda \end{bmatrix} \perp \begin{bmatrix} W_{nn} & W_{nt_1} & \ldots & W_{nt_k} & 0 \\ W_{t_1 n} & W_{t_1 t_1} & \ldots & W_{t_1 t_k} & 1 \\ \vdots & \vdots & \ddots & \vdots & \vdots \\ W_{t_k n} & W_{t_k t_1} & \ldots & W_{t_k t_k} & 1 \\ \mu & -1 & \ldots & -1 & 0 \end{bmatrix} \begin{bmatrix} f_n \\ f_{t_1} \\ \vdots \\ f_{t_k} \\ \lambda \end{bmatrix} + \begin{bmatrix} \delta_n^{\text{free}} \\ \delta_{t_1}^{\text{free}} \\ \vdots \\ \delta_{t_k}^{\text{free}} \\ 0 \end{bmatrix}. \quad (24)$$

Gathering the relations on each contact, the method then leads to a global LCP of size equal to $m \times (k+2)$ if $m$ is the number of contacts. To have sufficient precision $k = 8$ faces on pyramids are often used; this leads to an LCP 10 times bigger than the frictionless one. In the next section, a new approach that uses an iterative Gauss-Seidel-like algorithm is proposed.

## 5 GAUSS-SEIDEL-LIKE ALGORITHM

In the area of computational mechanics of granular simulations (see [26]), a Gauss-Seidel (GS) like algorithm has been proposed to solve Signorini's and Coulomb's laws with a guaranteed convergence. This method has been recently applied in robotics [27]. In this section, we first present an adaptation of this algorithm to deformable bodies; then, we compare its performance to previous $k$-sided pyramids in a real-time context.

### 5.1 Formulation

Considering a friction contact $\alpha$, among $m$ instantaneous contacts, one can write the linearized behavior of the system in the form:

$$\underbrace{\delta_\alpha - [W_{\alpha\alpha}]f_\alpha}_{\text{unknown}} = \underbrace{\sum_{\beta=1}^{\alpha-1}[W_{\alpha\beta}]f_\beta + \sum_{\beta=\alpha+1}^{m}[W_{\alpha\beta}]f_\beta}_{\text{frozen}} + \delta_\alpha^{\text{free}}, \quad (25)$$

where $[W_{\alpha\beta}]$ is a $(3 \times 3)$ matrix that models the coupling, in the Delassus operator, between the contacts $\alpha$ and $\beta$. The local solution must agree with Signorini's and Coulomb's laws.

The resolution method used can be compared to the block nonlinear Gauss-Seidel algorithm. A certain number of iterations are made contact by contact. On each contact $\alpha$, this method consists of solving the contact and friction laws by considering the contribution of other contacts ($\alpha \neq \beta$) "frozen." The solution of every block equations (contact, friction) is nonlinear and can be realized by a iterative method in 3D [28].

Some notable remarks are:

- Using the estimated value $\widetilde{f}_\alpha$, (25) can be rewritten in the form:

$$\delta_\alpha - [W_{\alpha\alpha}](f_\alpha - \widetilde{f}_\alpha) = \sum_{\beta=1}^{\alpha-1}[W_{\alpha\beta}]f_\beta + \sum_{\beta=\alpha}^{m}[w_{\alpha\beta}]f_\beta + \delta_\alpha^{\text{free}}.$$

- When we are near the solution, the value of $W_{\alpha\alpha}$ is not dominant because $f_\alpha - \widetilde{f}_\alpha \to 0$.

Hence, one may propose a method where $W_{\alpha\alpha}$ can be replaced by a diagonal matrix with values judiciously computed (as presented in Algorithm 1). Then, a method of graph intersection can solve the contact and friction laws. This method works well and is faster, even if it necessitates doing more iterations than using an iterative method to solve block equations.

---

**Algorithm 1:** Gauss-Seidel like resolution algorithm

**Input:** $(\delta^{\text{free}})_{(3m \times 1)}, [W]_{(3m \times 3m)}$
**Output:** $(f)_{(3m \times 1)}$
set $\epsilon_1$ to Signorini's law tolerance
set $\epsilon_2$ to desired precision
$k = 0$
**for** $i = 1 \ldots m$ **do**
 $(f_i^{(0)})_{(3 \times 1)} = 0$
 $\Lambda_{\text{min/max}} = \text{eig}([W_{ii}]_{tt})$
 $\Lambda_i = \frac{\Lambda_{\text{min}} + \Lambda_{\text{max}}}{2}$
**end**
**repeat**
 $k = k + 1$
 **foreach** $i = 1 \ldots m$ **do**
  $(\delta^{\text{test}})_{(3 \times 1)} = (\delta_i^{\text{free}})_{(3 \times 1)}$
  **foreach** $j = 1 \ldots i-1$ **do**
   $(\delta^{\text{test}})_{(3 \times 1)} += [W_{ij}]_{(3 \times 3)}(f_j^{(k)})_{(3 \times 1)}$
  **end**
  **foreach** $j = i \ldots m$ **do**
   $(\delta^{\text{test}})_{(3 \times 1)} += [W_{ij}]_{(3 \times 3)}(f_j^{(k-1)})_{(3 \times 1)}$
  **end**
  $(f_i^{(k)})_n = (f_i^{(k-1)})_n - (\delta^{\text{test}})_n/[W_{ii}](1,1)$
  **if** $(f_i^{(k)})_n > \epsilon_1$ **then**
   $(f_i^{(k)})_t = (f_i^{(k-1)})_t - (\delta^{\text{test}})_t/\Lambda_i$
   **if** $\|(f_i^{(k)})_t\| > \mu(f_i^{(k)})_n$ **then**
    $(f_i^{(k)})_t \times = \frac{\mu(f_i^{(k)})_n}{\|(f_i^{(k)})_t\|}$
   **end**
  **else**
   $(f_i^{(k)}) = 0$
  **end**
 **end**
**until** $\sum_{i=1}^{m} \frac{\|f_i^{(k)} - f_i^{(k-1)}\|}{\|f_i^{(k)}\|} < \epsilon_2$;

---

### 5.2 Performance

In this section, we compare the GS-like approach to the LCP one (i.e., when $k$-sided pyramids are used). The test simulates a small deformable table, with four legs, that is in contact with the floor (see Fig. 6). All simulations have been performed using MATLAB©. As results are biased by implementation issues, figures mainly show a tendency.[8]

First, we can compare the two approaches in terms of their precision. Vector $f_{\text{LCP}}$, the outcome of the pyramid approach, is compared to the vector $f_{\text{GS}}$ computed by the GS-like algorithm. The formula used to measure the gap $\gamma$ between the two vectors is:

$$\gamma(\%) = 100 \times \frac{\|f_{\text{LCP}} - f_{\text{GS}}\|}{\|f_{\text{GS}}\|}. \quad (26)$$

---

8. The asymptotic complexity of the $k$-sided pyramids algorithm is $\mathcal{O}(k \times m^2)$, where the asymptotic complexity of the Gauss-Seidel algorithm is, in the worst case, $\mathcal{O}(m^2)$ (see Section 5.4).

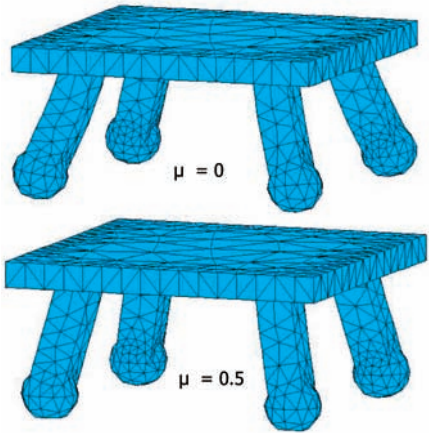

Fig. 6. The same table with the same load. This example shows the importance of the friction coefficient to the behavior of the table.

Tests have been performed with a small friction coefficient ($\mu = 0.1$) to have dynamic friction (see Fig. 7).

As the GS approach does not introduce any approximation on the friction cone, the precision found with this algorithm is better than with the LCP one. The test shows that, to have $\gamma < 5$ percent between the two approaches, it is necessary to use 16-sided pyramids.

However, a more significant issue is the CPU processing time when introducing dry friction between deformable models with haptic feedback. Thus, in Fig. 8, we have compared the computing time in 30 different cases of a load on the table.

These tests clearly show that our proposed GS approach is much more efficient when the number of instantaneous contacts increases. We also demonstrate that, on the contrary to what we would presuppose, the use of $k$-sided approximations, even if they keep having LCP, should be called into question in computer haptics; see Section 7 for a thorough discussion about performance.

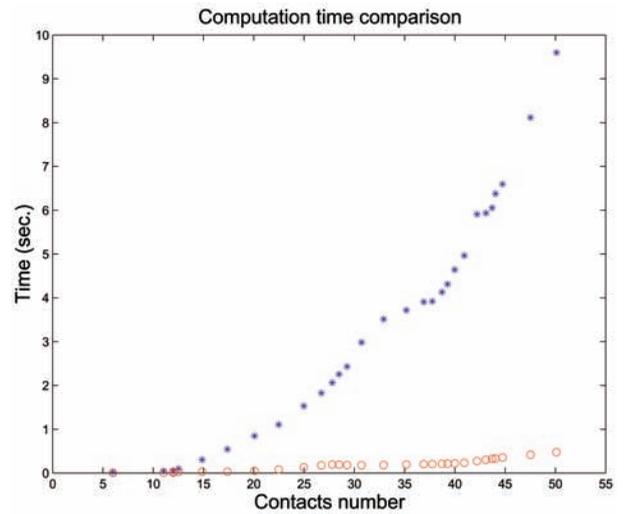

Fig. 8. Average CPU time, obtained from 50 tests with random motion in tangential plane and random friction coefficient. Red circles denote nonlinear friction cone resolution by GS, whereas the blue stars denote the LCP resolution of the same problem using 8-sided pyramids.

### 5.3 Mechanical Coupling between Contacts

By introducing the Delassus operator $[W]$, computing the contact and friction force takes account of the material and the structural property of the contacts. As in the multicontact case, each contact may be linked to others. The coupling values are taken into account in the process through Delassus operator. That makes a difference compared to penalty methods. Also, the existence of couplings is the reason why single contact methods cannot be adapted to multicontacts easily.

To illustrate this, we have created a *visualization* for the Delassus operator (Fig. 9) of the virtual table in contact with floor. It is obvious that two contacts on the same leg are much more coupled than two contacts on different legs. Thus, four blocks of contacts appear, corresponding to the contact of the four legs.

The coupling between contacts on different legs is very small. By using reinforcements between the legs (Fig. 10),

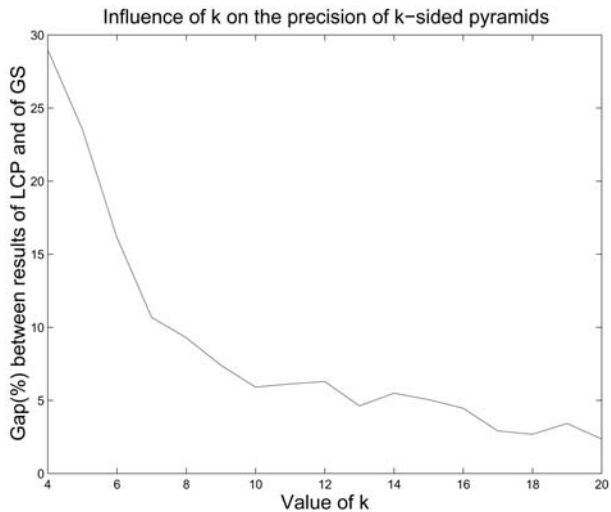

Fig. 7. Average gap $\gamma$ over 30 tests where a random motion in the tangential plane is introduced to produce different slip directions.

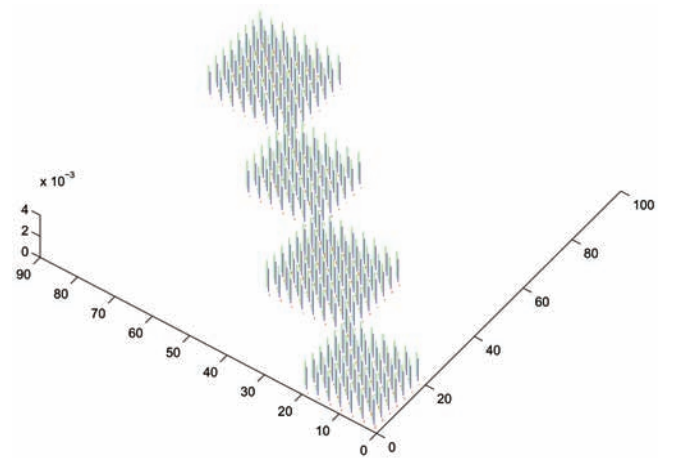

Fig. 9. Visualization of the Delassus operator during the contact between the virtual table and the floor. In red, couplings are along the normal and in blue and green, they are along tangential directions.

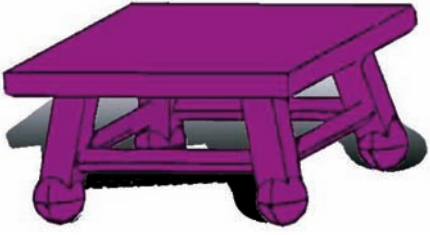

Fig. 10. Table with reinforcements.

we can increase the rigidity of the structure and also the coupling between contacts.

On one hand, the tangential values of $[W]$ are lower. Indeed, augmentation of stiffness induces smaller compliance. On the other hand, Fig. 11 shows that the coupling between contacts on different legs increases along the direction of reinforcements.

Our process includes the mechanical properties of multicontacts by using the Delassus operator which reflects the linearized behavior of bodies in the contact space. Thus, it will allow us to solve contacts and friction between models that do not have the same mechanical impedance.

However, in dynamics, this computation may induce constraints on the choice of time step. A solution to this problem is presented in the following section.

### 5.4 Scalability of the Algorithm

The proposed algorithm deals with contact and friction resolution. Therefore, its complexity depends on the number of contacts and their coupling, and not on the shape or the number of nodes of the object which are more problematic for collision/proximity algorithms.

Since the Gauss-Seidel algorithm for frictional contacts is based on iterative computations, it will output only approximate solutions. The total number of operations depends on the convergence velocity and on the tolerance (i.e., $\epsilon_2$) chosen to stop the algorithm. Note that this tolerance may be a useful tool for tuning between computation time and precision.

As it is known for all Gauss-Seidel methods, the convergence velocity depends on the dominance of diagonal values. This property of diagonal dominance in the Delassus operator is verified with deformable models that use a finite element method in 3D linear elasticity.

If the deformation law is chosen to be linear (small displacement case), a lot of computation need only be made once. For instance, $\tilde{K}$ can be condensed on the surface nodes and factorized in a precomputed process. Then, to compute the Delassus operator, one need only transform the frames of reference (from the object's frame to the contact frame) and interpolate (from the object's node to the contact point). The complexity of this operation is $\mathcal{O}(m^2)$, $m$ being the number of contacts, if there is mechanical coupling between all contacts (in the worst case).

Moreover, if this computation of the Delassus operator is too long, one can also build only the block diagonal part of the matrix and calculate the frozen contribution of each contact (see the right side of (25)) in each model frame of reference (using $\tilde{K}^{-1}H^T F$) and map it to the contact frame.

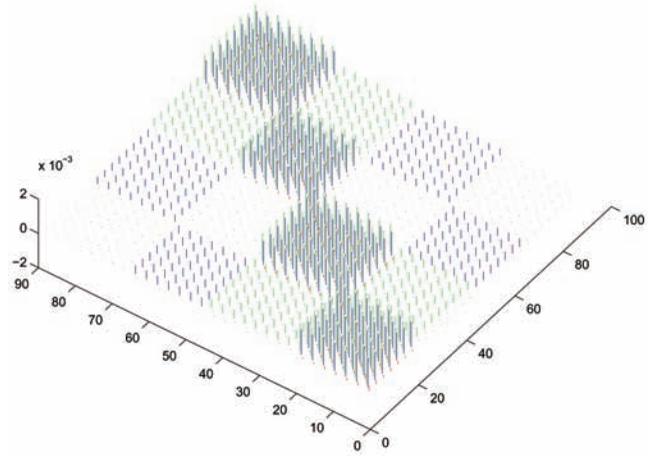

Fig. 11. The Delassus operator of the same table with reinforcements (along one tangential direction, in green, along the other direction, in blue).

The main computation during one iteration concerns the contribution of all frozen contacts to the computed contact. This computation is equivalent to a matrix-vector product for each iteration. So, one iteration has an $\mathcal{O}(m \times m')$ complexity where $m'$ is the number of active contacts (whose force is not null) if the Delassus operator is full, and fewer operations if the operator is sparse. This is also known to be the complexity of the GS algorithm.

## 6 COROTATIONAL APPROACH

Haptic rendering for rigid body simulation requires update rates ranging from 500Hz to 1kHz. As our simulation runs in real-time, the time step used in the simulation is about one or two milliseconds. In this section, the dynamic case is considered. The choice of the time step depends on the mass/stiffness ratio of the deformable objects. This may compromise the simulation of deformable light and structured materials. A global corotational approach that decouples a rigid global motion from a deformable local one is incorporated. This method allows larger tolerance on the time step which consequently allows haptic feedback to be performed on all kinds of material, even very stiff ones, without modification of the simulation parameters.

### 6.1 Time Step Choice

Consider (1) in its linearized form. Applying the Euler implicit scheme leads to the following equation:

$$\underbrace{\left(\frac{[M]}{\Delta t^2}+\frac{[D]}{\Delta t}+[K]\right)}_{\tilde{K}} \mathbf{u}_t = \underbrace{\mathbf{f}_t+\left(\frac{[M]}{\Delta t^2}+\frac{[D]}{\Delta t}\right)\mathbf{u}_{t-1}+\frac{[M]}{\Delta t}\dot{\mathbf{u}}_{t-1}}_{\mathbf{f}_{(t,t-1)}}. \quad (27)$$

If no Dirichlet conditions are defined, the stiffness matrix $[K]$ is singular. In this case, the value of $\Delta t$ must be adapted in order to make, in $\tilde{K}$, the mass and damping matrices *dominant* compared to the stiffness matrix.

Even if such conditions are defined, the problem is not completely solved. If one increases $[K]$ without decreasing $\Delta t$, the deformation frequencies would not satisfy the well-known Shannon theorem. Consequently, there is no other solution than adapting the time step, according to the "ratio" between the mass and the stiffness of body. In this case, since the limit of the time step is known, we can adapt the Gauss-Seidel-like iteration to satisfy the time step by relaxing $\epsilon_2$.

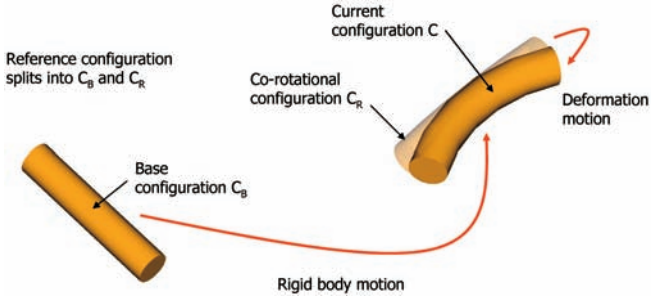

Fig. 12. Adapted from Felippa [30]. The motion of a deformable object may be split in two parts: a deformable motion in its current configuration and a rigid motion in the space.

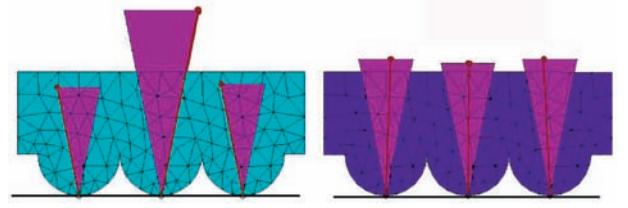

Fig. 13. A rigid model in frictional contact with the floor. Friction cones are represented by purple triangles and contact forces by brown arrows. We show the solutions given by the Gauss-Seidel algorithm when only the rigid model is considered (on the left) and when the corotational global approach is used (on the right). Both solutions result in the same behavior (the rigid body does not move), but with our method the contact forces are much more realistic, from the physical viewpoint, and there is only one solution.

## 6.2 Delassus Operator with a Corotational Global Approach

The approach proposed in [29] describes the motion of deformable bodies. This model splits the global motion (driven by a rigid model) from local relative displacement (driven by a linear deformable model), as shown in Fig. 12. Recent developments in corotational approaches can be found in [30] and in [31]. We are using a global corotational approach, which differs from the local one that might be used for large deformations, see [32].

The rigid dynamic models can be written in their generalized form (for details, see, for instance, [12]):

$$A(q)\ddot{q} + b(q,\dot{q}) = \Gamma^{\text{external}} + \Gamma^{\text{constraint}}, \quad (28)$$

where $A(q)$ is a mass and inertia matrix, $q$ is the vector of generalized degrees of freedom, $b(q,\dot{q})$ the centrifugal and Coriolis force vector, $\Gamma^{\text{external}}$ is the resultant of external forces, and $\Gamma^{\text{constraint}}$ is the resultant of contact forces.

The methodology proposed in [12] is used to add the rigid motion in the contact space. In this work, a Jacobian $J_c$ is introduced to map the motion space into the contact space (for instance, the resultant $\Gamma^{\text{constraint}}$ of contact forces $F$ is computed with $\Gamma^{\text{constraint}} = J_c^T F$). This allows us to obtain the acceleration of the gaps in the contact space due to rigid motion:

$$\ddot{\delta}^{\text{rigid}} = [J_c A^{-1} J_c^T] f + \ddot{\delta}^{\text{free}}. \quad (29)$$

With the corotational global approach, the motion of one point is the sum of its rigid motion in the global space and its local deformation (3). Thus, the two models can be summed into compliance within the contact space. The linearized behavior of the models may be written as:

$$\delta = \underbrace{\left[\sum_i H_i C_i H_i^T + J_c\left(\frac{A}{\Delta t^2}\right)^{-1} J_c^T\right]}_{[W]} f + \delta^{\text{free}}. \quad (30)$$

With this model, when the stiffness of the body increases, the behavior tends to rigid body motion. And, in the Delassus operator $[W]$, mass and stiffness are decoupled. This is why this model allows stable haptic feedback and real-time simulation with an arbitrary choice of time step.

As an additional justification, we recall that modeling a body as rigid or as deformable is not a matter of the object's intrinsic properties. Indeed, the model depends rather on the ways each object is constrained and the nature of these constraints. In most cases, a rigid model is a valid simplification only when the deformations can be neglected.

## 6.3 "Quasi-Rigid" Application

Using a corotational global model, our method includes frictional rigid contacts in the limit of increasing stiffness. In this case, the deformable part can only be seen as a physically plausible mechanical compliance that solves part of the indetermination that appears when Coulomb's friction law is adopted.

Indeed, it is known that, in rigid body mechanics, frictional extensions lead to nonunique solutions [23], [25] that, however, respect Coulomb's law and result in the same rigid motion, as shown in Fig. 13. These nonuniquenesses usually induce convergence problems. The nonlinear Gauss-Seidel method is able to obtain one result even if there is more than one solution, but the solution is influenced by the contact treatment ordering.

Very recent work [33], [34] has proposed adding small deformations to rigid objects in order to add sufficient degrees of freedom and to obtain solutions which are always unique and smoother. Our approach can be used in exactly the same way. However, what distinguishes our method from previous ones is that we use a mechanical compliance based on an FEM model, leading to more plausible results.

## 7 FORCE FEEDBACK COMPUTATION AND STABLE HAPTIC FEEDBACK ISSUES

The models and algorithms we described have been implemented, packaged, and experimented with haptic feedback scenarios. First, the coupling used to allow haptic feedback on a corotational model is described. Then, snap-in tasks with computations timing results are presented together with our haptic set-up.

### 7.1 Force Feedback Coupling

Our approach here is a continuation of Adams and Hannaford's work [35], where the stability is straightfor-ward and the feedback forces depend strongly on the behavior of the object. However, the update rate will influence the parameters used for the coupling.

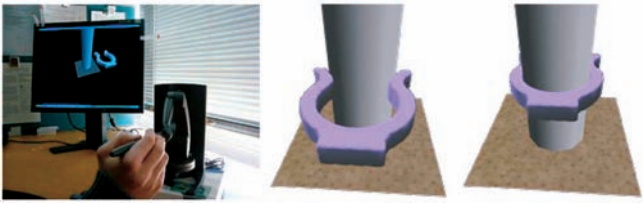

Fig. 14. Haptic feedback simulation of a snap-in operation with a flexible tool into a "quasi-rigid" pipe.

Indeed, a 6D virtual coupling between the interface and the rigid part of the corotational model uses impedance for the interface and admittance for the simulation. This virtual coupling may be considered intuitively as a 6 degrees of freedom stiffness and damping between positions and velocities measured at the interface and given by the simulation. This creates the force that is sent both to haptic feedback and to the real-time simulation. To obtain intuitive deformations, Dirichlet conditions for the deformable part of the model are defined in the neighborhood of the grabbed spot.

### 7.2 Snap-In Task

The example we give in this paper is a virtual snap-in task between two objects, one being deformable (a clip) and the other being either rigid or deformable (a pipe). This example has been chosen (actually proposed by our collaboration with industry in the field of virtual prototyping (VP) development), because it has a nonlinear behavior from the computer haptics point-of-view, and requires a sustained haptic perception/action coordination.

To demonstrate the device independence given by the adopted methodology, we implemented the similar scenario using two different setups. The first configuration uses an ordinary PC with a PHANToM Desktop from Sensable as a haptic device (Fig. 14).

The second setup uses two haptic 6dof devices, Virtuose6D, from Haption and a workbench from BARCO. This configuration offers the possibility to blend in the same space haptic and visual rendering and allow the operator to work in a standing position, see Fig. 15.

The scenario is defined as follows: First, the operator needs to grab the virtual clip, and to move it to the pipe where it should be snapped-in. Through the haptic device, the operator reaches for the clip's handle, and by a simple button press, attaches it to the haptic device by a virtual 6 degrees of freedom coupling. When the clip reaches the pipe spot, the first collisions occur; subsequent haptic feedback will help the operator to correctly position the clip on the pipe to start the snap-in process. The snap-in process consists of three phases: a pushing phase, an unstable equilibrium phase, and the final clipping phase.

The pushing phase consists of the operator applying forces on the clip toward the pipe. Here, the contact points stick due to static friction. When the deformation starts, there is an induced resistance due to deformation forces and dynamic friction, until the deformation reaches its max-imum (the distance between the two branches of the clip is maximum). At this moment, we reach the second phase.

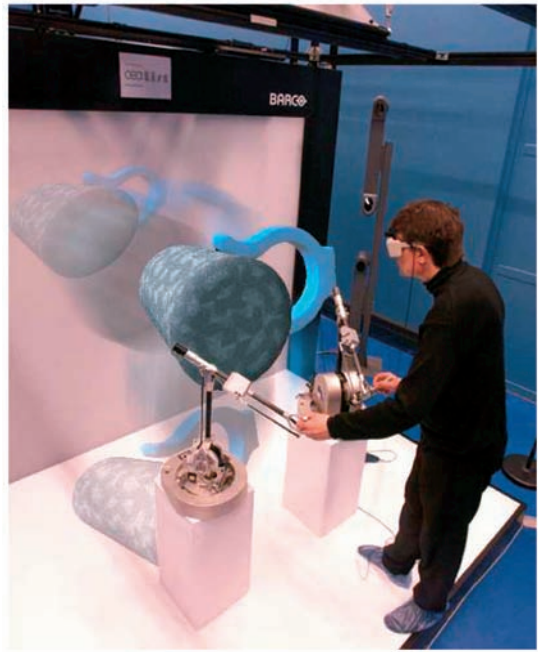

Fig. 15. CEA/LIST virtual prototyping configuration set-up. Here, two 6 degrees of freedom force feedback devices VIRTUOSE6D are used.

The second phase is instable since it is an instant-time state. At this instant, if the applied forces decrease, the clip may come back in an abrupt way, especially if the object to be snapped-in is rigid, as in Fig. 14. If the forces are sufficient, the clip goes to the third phase.

In the third phase, the motion of the clip is relatively abrupt, especially if the pipe is rigid, since the closing clip forces induced from the deformation relaxation are significant.

Fig. 16 shows screen snapshots from deformable/deformable interactive snap-in. When both objects are deformable, it is not easy to keep coordination of the motion and force. The most impressive haptic sensation is when user tries to withdraw the clip. Here, static friction and clip deformation give a significant resistance to the motion of the clip, while, as the user moves the interface, the haptic virtual coupling accumulates potential energy. Thus, in the very last phase, the clip moves out from the cylinders in a very abrupt way.

As this motion could be very fast, this example shows the importance of the implicit resolutions of models, contact, and friction laws. The values used to calculate the stiffness of the clip (Young Modulus $E = 700 MPa$ and Poisson coefficient $\mu = 0.35$) correspond to polyethylene. The corotational method allows us to use a realistic mass (15g) with a time step of 3ms.

### 7.3 Performance

The snap-in demos use a stochastic proximity detection algorithm. We can fix the number of the shortest distances given at each time step (close to what is presented in [10]).

Fig. 17 illustrates the small variations of the computation time when solving thirty contacts without friction between a pair of objects. The mean computing time is about 3ms. Contact computations allow this very constrained case to be solved with truthful behavior.

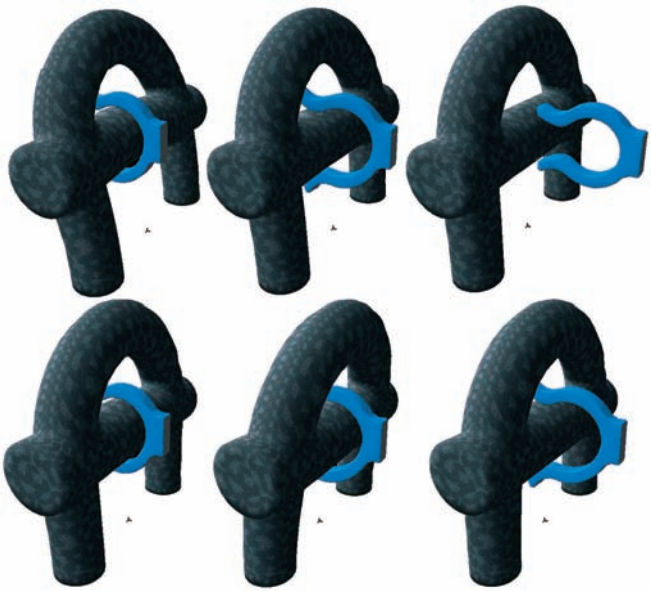

Fig. 16. Snapshots of the interactive snap-in and snap-out task on deformable pipes. At the top, the pushing phase of the snap-in task. At the bottom, the user withdraws the clip. This creates deformations of the pipes, especially if the friction coefficient is very large.

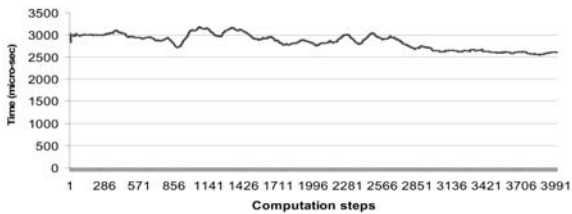

Fig. 17. CPU time for treatment of 30 instantaneous frictionless contacts.

When friction is added to the simulation, the performance of course decreases. This is shown in Fig. 18. We obtain a CPU time of 4ms with 20 instantaneous proximity spots. However, virtual coupling allows us to still have stable haptic feedback with a relatively low refresh frequency (up to 250Hz).

## 8 CONCLUSION AND FUTURE WORK

When flexible virtual objects are interactively manipulated, stable and truthful computer haptics requires physical modeling of contact and friction in addition to real-time force and deformation computations. The originality of this work is in formulating an implicit solution that solves Signorini's and Coulomb's laws in a very fast manner, thanks to a Gauss-Seidel-like algorithm. The proposed solution can be combined with most existing fast deformable simulations, as we disassociate the contact treatments from the deformation behavior. We also show the interest of using a global corotational approach in a real-time simulation context. Our solutions have been implemented and tested; a force feedback virtual snap-in simulation is described and its performance evaluated.

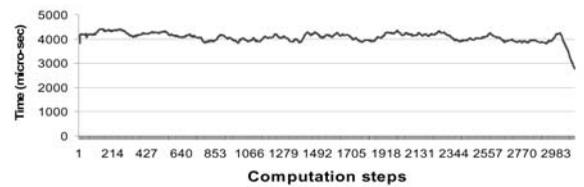
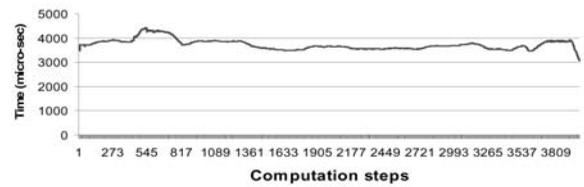

Fig. 18. CPU time for the treatment of 20 instantaneous contacts with (a) $\mu = 0.2$ and (b) $\mu = 0.8$.

Our future work will focus on performance optimization of the computation process and the investigation of other dry friction models. We also plan to apply these techniques in a medical simulation context.

## ACKNOWLEDGMENTS

This work was supported by the CEA.

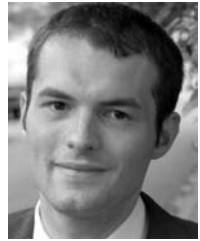
**Christian Duriez** received the engineering degree from the Institut Catholique d'Arts et Métiers of Lille, France, a master of science degree by research, and the PhD degree in robotics, both from University of Evry, France. His thesis work was realized at CEA/Robotics and Interactive Systems Technologies. The primary topic of his current postdoctoral position at CIMIT/Harvard Medical School is how to include contact and friction in realistic medical simulations. He is a student member of the IEEE.

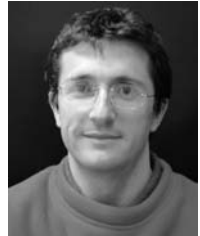
**Frédéric Dubois** graduated from the Ecole Supérieure d'Ingénieurs de Marseille in 1990. He received the PhD degree in mechanics from the University of Aix-Marseille 2 in 1994. He is now a CNRS research engineer in the Laboratoire de Mécanique et Génie Civil of Montpellier. His research activity mainly includes numerical modelling, numerical methods, and intensive computation. He manages the development of an open source software (LMGC90) devoted to the simulation of multibody contact problems.

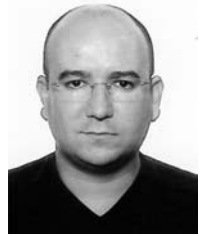
**Abderrahmane Kheddar** received the computer science engineering degree from the Institut National d'Informatique (INI) Algiers, Algeria, a DEA (Master of Science by research), and the PhD degree in robotics, both from the University of Paris 6, France. He is currently a professor at the University of Evry, France. He is the head of the virtual reality and haptics team in LSC (CNRS) and, in the frame of his CNRS secondment, he is the codirector of the JRL in Tsukuba, Japan. His research interests include haptics, virtual reality, and humanoids. He is a member of the IEEE.

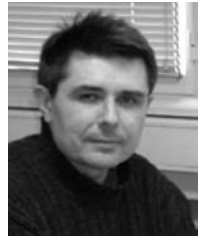
**Claude Andriot** received the PhD degree in robotics from the University of Paris 6 in 1992. He is currently at CEA (French nuclear authority) in the Robotics and Interactive Systems Technologies Lab. He is the CEA project leader of the Perfrv platform (French Platform on Virtual Reality, http://www.perfrv.org). His research activities and projects involve control of force feedback devices for desktop and nuclear applications, control of wearable haptic interfaces for virtual reality, and real time physical simulation of mechanical systems.